\documentclass[12p]{paper}
\usepackage{amsmath}
\usepackage{amssymb}
\usepackage{authblk}
\usepackage{graphicx}

\begin{document}

\title{Response of Windowless Silicon Avalanche Photo-Diodes to Electrons in the 90-900~eV Range}

\author[1]{Alice Apponi}
\author[1]{Gianluca Cavoto}
\author[2]{Marco Iannone}
\author[1]{Carlo Mariani}
\author[2,*]{Francesco Pandolfi}
\author[3]{Daniele Paoloni}
\author[2]{Ilaria Rago}
\author[4]{Alessandro Ruocco}

\affil[1]{`Sapienza' Universit\`{a} di Roma e INFN Sezione di Roma, Piazzale Aldo Moro 2, 00185 Rome, Italy}
\affil[2]{INFN Sezione di Roma, Piazzale Aldo Moro 2, 00185 Rome, Italy}
\affil[3]{Dipartimento di Scienze Universit\`a degli Studi Roma Tre, Via della Vasca Navale 84, 00146 Rome, Italy}
\affil[4]{Dipartimento di Scienze Universit\`a degli Studi Roma Tre, and INFN Sezione di Roma Tre, Via della Vasca Navale 84, 00146 Rome, Italy}
\affil[*]{Corresponding author. E-mail address: francesco.pandolfi@roma1.infn.it}

\maketitle

\abstract{We report on the characterization of the response of windowless silicon avalanche photo-diodes to electrons in the 90-900~eV energy range. The electrons were provided by a monoenergetic electron gun present in the LASEC laboratories of University of Roma Tre. We find that the avalanche photo-diode generates a current proportional to the current of electrons hitting its active surface. The gain is found to depend on the electron energy~$E_e$, and varies from $2.147 \pm 0.027$ (for $E_e = 90$~eV) to $385.8 \pm 3.3$~(for $E_e = 900$~eV), when operating the diode at a bias of $V_{apd} = 350$~V.} This is the first time silicon avalanche photo-diodes are employed to measure electrons with $E_e < 1$~keV.

\newpage
\clearpage

\section{Introduction}

Silicon avalanche photo-diodes (APDs) are widely employed as photon detectors, however they can also be used to detect electrons with energy $E_e \lesssim 100$~keV. In particular, the use of APDs to detect electrons in the medium energy range~($10-100$~keV) has been studied quite extensively in recent years, in particular for applications in space missions~\cite{space0,space1,space2}, where the APD durability, combined with its insensitivity to magnetic fields, are attractive features. While some studies have been performed to use APDs to detect lower-energy charged particles~\cite{ogas}, the use of APDs to detect low ($<1$~keV) energy electrons is a less studied field, and is the topic of this work.

\begin{figure}[hbt]
  \centering
  \includegraphics[width=0.8\textwidth]{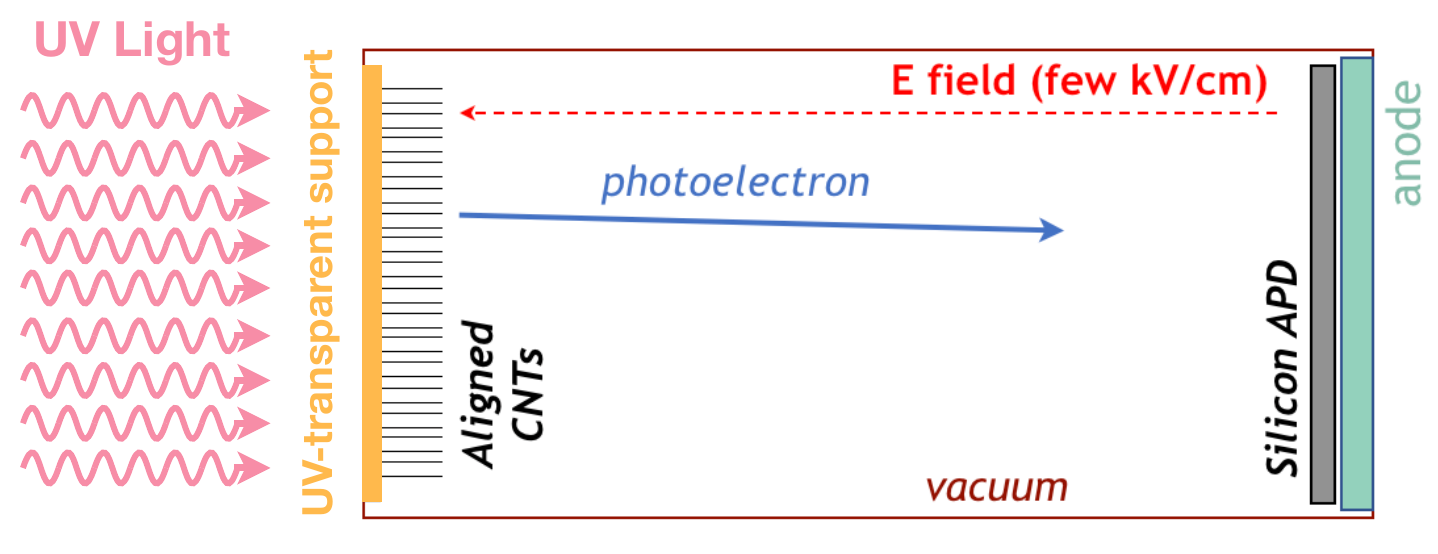}
 \caption{Schematic view of the NanoUV detector concept.
  \label{fig:nanouv}}
\end{figure}

The results presented in this paper are produced in the context of the development of a novel UV light detector (NanoUV), with a photocathode made of vertically-aligned carbon nanotubes~\cite{cavoto0, cavoto, antochi, cavoto2}. Vertically-aligned carbon nanotubes can be grown with chemical vapor deposition techniques~\cite{rago} up to lengths of a few hundreds of $\mu$m, with the result of obtaining a highly anisotropical material, with ideally vanishing density in the tube axis direction~\cite{mariani,mariani2}. A photocathode made of such material could significantly reduce the probability of photo-electron re-absorbtion, which is the leading cause of inefficiency for modern day UV-light detectors, because photoelectrons would be emitted directly into the vacuum, and would be able to exit the nanotubes if their momentum is parallel to the tubes. The electrons are then accelerated by an applied potential $\Delta V \lesssim 10$~kV, and then detected by a silicon APD placed at the other end of a vacuum tube a few centimeters long. A schematic view of the NanoUV detector concept can be seen in Figure~\ref{fig:nanouv}.

\begin{figure}[htb]
  \centering
  \includegraphics[width=0.47\textwidth]{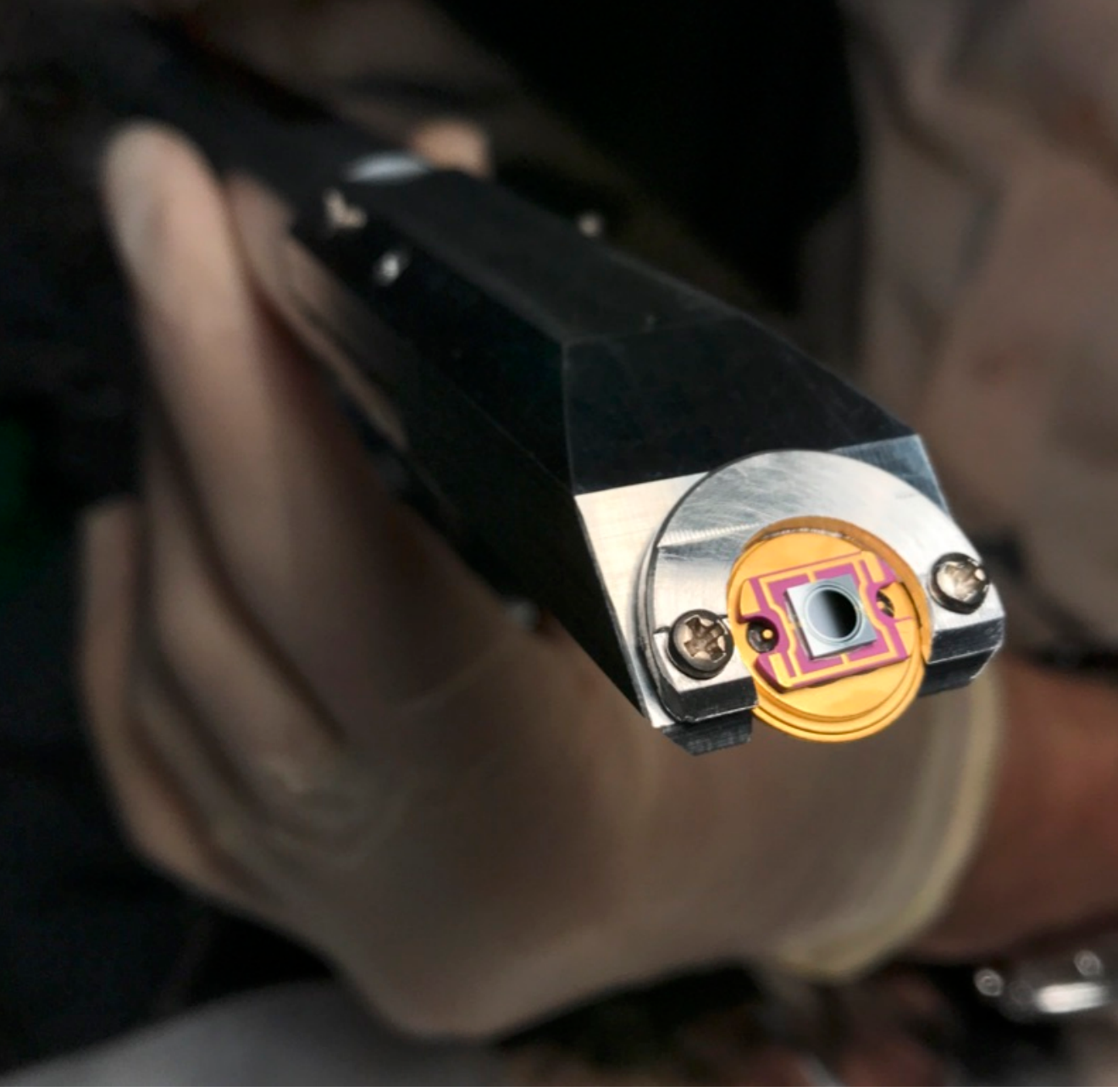}
 \hspace{0.5cm}
  \includegraphics[width=0.46\textwidth]{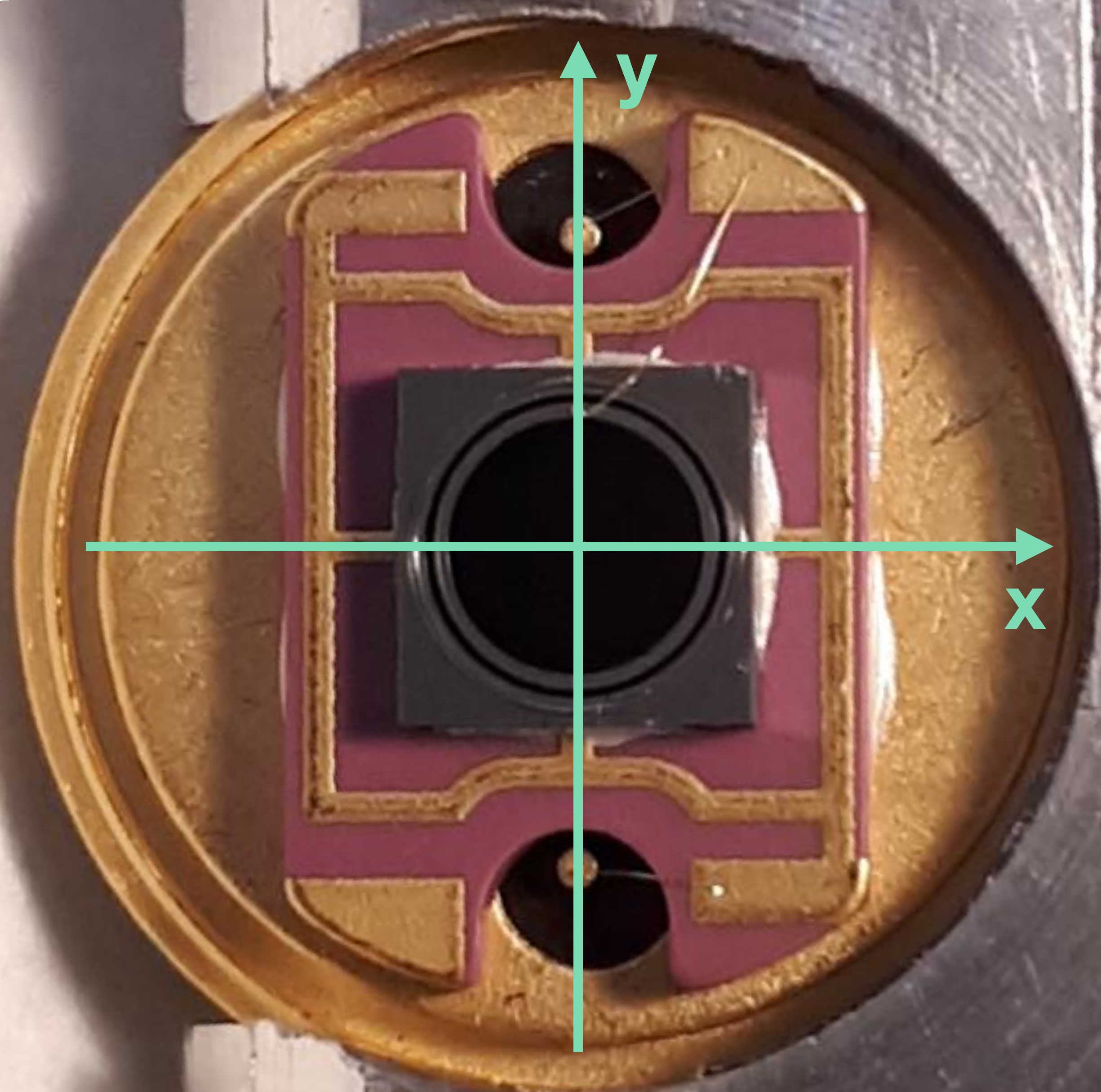}
   \caption{Left: avalanche photo-diode S11625-30N by Hamamatsu Photonics mounted on the custom-made support to be inserted in the UHV chamber. Right: detail of the front face of the diode, and definition of the $x$-$y$ reference system used in this paper.
  \label{fig:apd}}
\end{figure}

\section{Experimental Apparatus}

Commercial APDs typically cover the silicon sensor with a protective layer (or `window'), which is transparent for photons, but would result in the absorption of low-energy electrons, thus compromising its performance as an electron detector. For this reason we are employing special window-less APDs manufactured by Hamamatsu Photonics (S11625-30N), which can be seen in Figure~\ref{fig:apd}: the left picture shows the APD mounted on its custom-made support, right before inserting it in the UHV chamber; the right picture shows a detail of the APD front face, together with the $x$-$y$ reference system used in this paper. The active area of the APD is circular, and has a diameter of 3~mm. According to the factory specifications, the APD we tested has a gain $G_{\gamma} = 50$ for photons with $\lambda = 650$~nm, when operating at a bias $V_{apd} = 355.5$~V and at a temperature $T = 25^{\circ}$C. The APD gain depends on the mean free path of the electrons, and therefore on temperature~\cite{cms}. While the temperature of the laboratory was kept stable at $(23\pm1)^{\circ}$C, no specific cooling was employed on the device itself, therefore small temperature-dependent drifts in the APD response can be expected as a result of heating during operation.

\begin{figure}[tb]
  \centering
\includegraphics[width=0.69\textwidth]{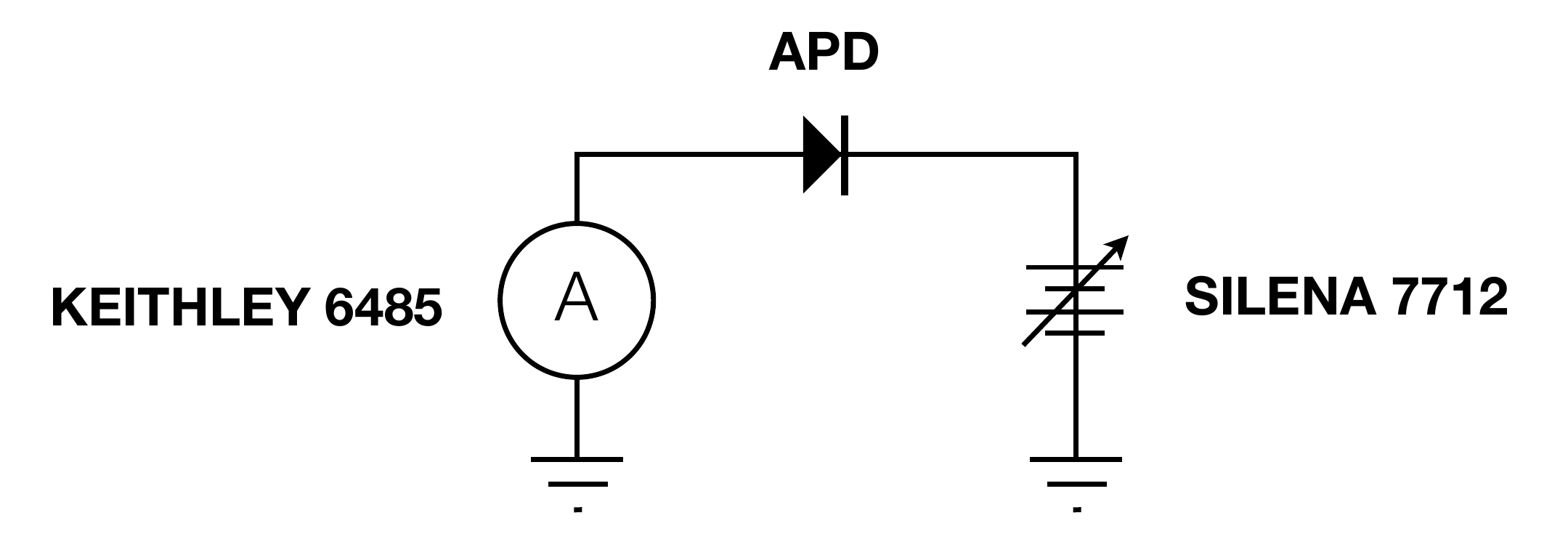}
 \caption{Schematic diagram of the operation and read-out of the APD.
  \label{fig:schema}}
\end{figure}

The APD is powered by a Silena 7712 power supply, at a nominal working bias of $V_{apd} = 350$~V. Throughout this paper all APD measurements are performed with a Keithley 6485 picoammeter connected to the ground pin of the APD, as shown in the diagram  in Figure~\ref{fig:schema}. 

\begin{figure}[tb]
  \centering
\includegraphics[width=0.69\textwidth]{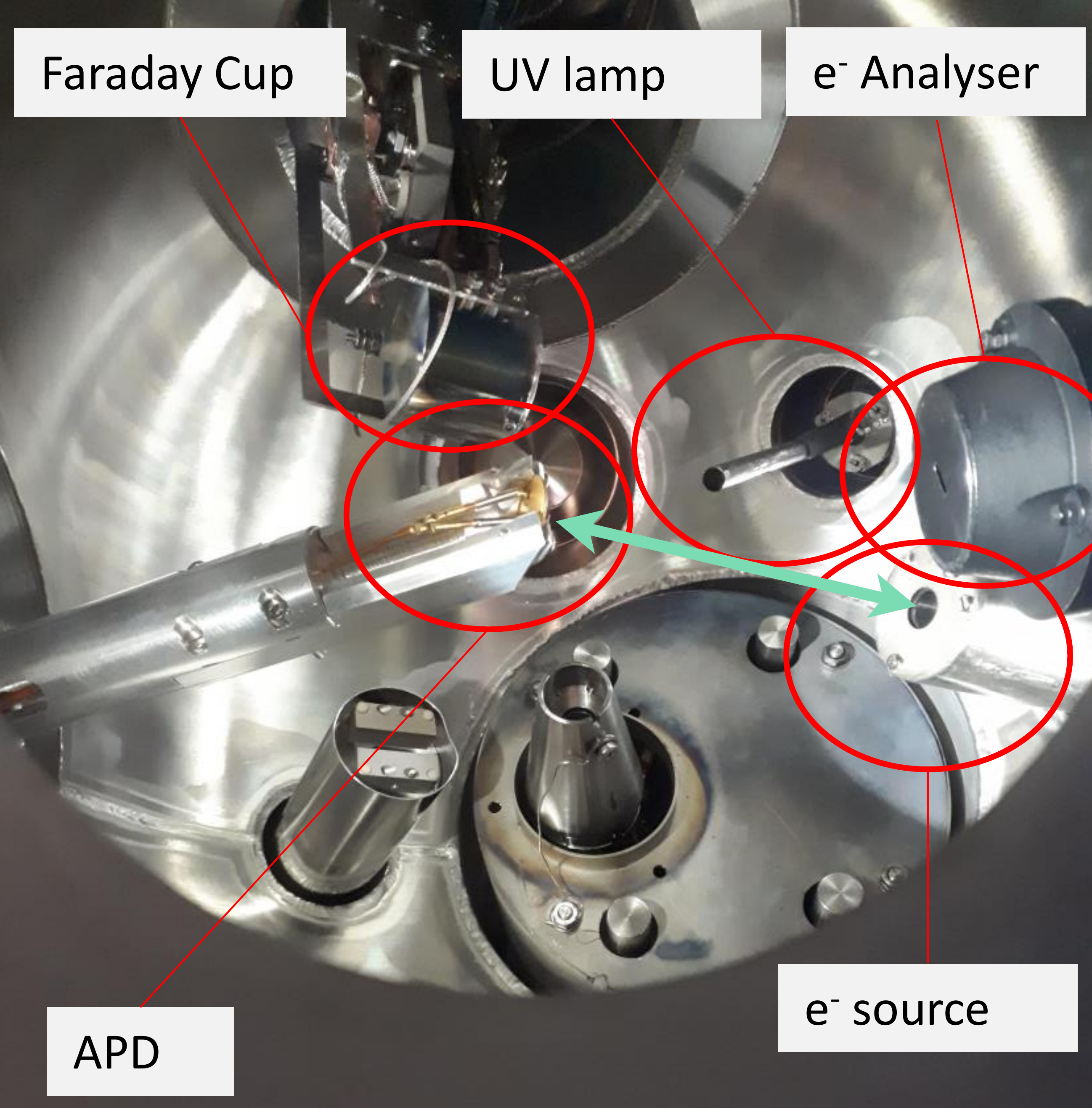}
 \caption{Picture of the interior of the UHV chamber present in LASEC labs in University of Roma Tre: the electron gun (`e$^{-}$ source') shoots electrons towards the center of the chamber~(green arrow), where either the APD or the Faraday cup can be present. The electron analyzer, used in the XPS analysis, is also shown, while the X-ray source is on the top of the chamber, outside of the picture frame.
  \label{fig:gun}}
\end{figure}

The APD characterization was performed in the LASEC laboratories at Roma Tre University. The APD is mounted on a custom-made anticorodal support, and is inserted inside the UHV chamber, where it can be brought to the line of sight of the electron gun. Alternatively, when the APD is kept in its retracted position, a Faraday cup for the precise measurement of the beam current and profile can be brought in the gun focus region. A schematic view of the apparatus can be seen in Fig.~\ref{fig:gun}.

The electron gun comprises a hot tungsten filament, followed by a system of electrostatic lenses. The energy spread is reduced by an electron monochromator based on two concentric hemispheres~\cite{kuyatt}. Additional metal plates are used to deflect the electron beam in the $x$ and $y$ directions. The gun is capable of producing beams of electrons between about 90 and 900~eV, with an energy dispersion of less than 0.05~eV. The electron beam current can be as low as a few fA and is measured through a Keysight B2987A picoammeter (nominal resolution: 0.01~fA).

\section{Experimental Procedure}

Windowless APDs have a superficial layer of silicon dioxide (SiO$_2$), which is typically grown during the manufacturing process. Energy deposited in the oxide layer will not produce a signal, so it is {\em de facto} a `dead' layer which needs to be crossed by the incoming electrons for them to be able to create a signal in the sensitive region of the silicon. It is therefore important to measure the thickness of the oxide layer.

\begin{figure}[tb]
  \centering
\includegraphics[width=0.69\textwidth]{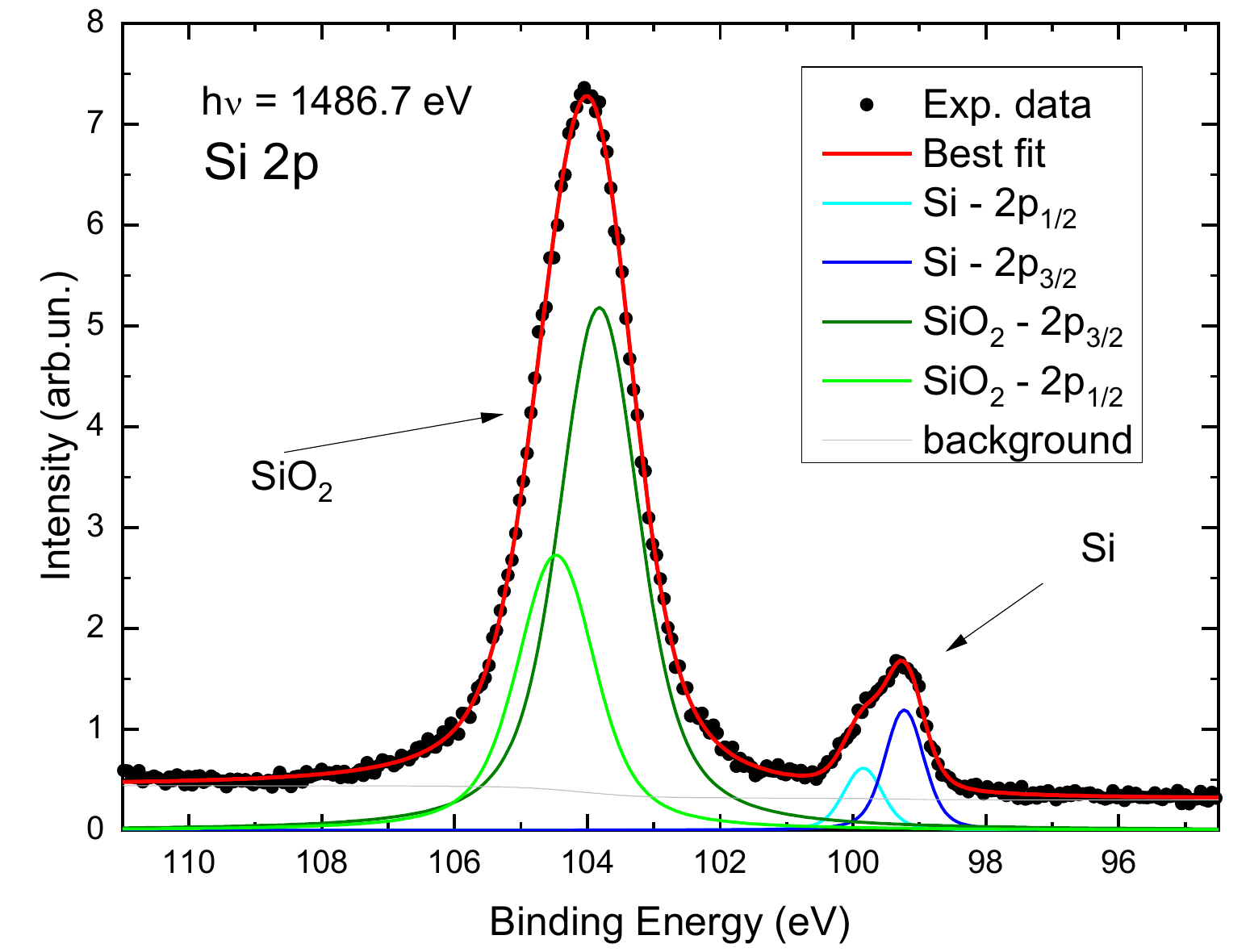}
 \caption{Binding energy of the photoelectrons ejected by the APD silicon during Al K$\alpha$ XPS. The spectrum is fitted with the expected shapes for 2p$_{1/2}$ and 2p$_{3/2}$, both for SiO$_2$~(shades of green) and for Si~(shades of blue). The background contribution is shown in grey. 
  \label{fig:oxideLayer}}
\end{figure}

This has been done with an X-ray photoelectron spectroscopy (XPS) analysis~\cite{xps}. X-rays from a Omicron XM1000 monochromatized Al K$\alpha$ source ($h\nu = 1486.7$~eV) are directed on the APD surface, and the ejected photoelectrons are detected by the electron analyzer (see Fig.~\ref{fig:gun} for their positioning inside the UHV chamber). Details of the experimental apparatus can be found here \cite{ruocco}. Knowing the photon energy and measuring the electron energy, one can infer the binding energy of the ejected electrons. This is shown in Figure~\ref{fig:oxideLayer}, which is centered around the typical values of the Si/SiO$_2$ 2p-orbitals. One can clearly see the peak corresponding to SiO$_2$, and also the contribution of the Si layer underneath it. Each peak is fitted with two Voigtian profiles in order to take into account the spin-orbit splitting between the 2p$_{1/2}$ and 2p$_{3/2}$ components. From the relative intensity of Si compared to SiO$_2$, taking into account the inelastic mean free path of the photoemitted electrons, one can estimate the depth of the Si layer, or, in other words, the SiO$_2$ thickness. We find this to be $8.5 \pm 1.0$~nm.

\begin{figure}[tb]
  \centering
\includegraphics[width=0.99\textwidth]{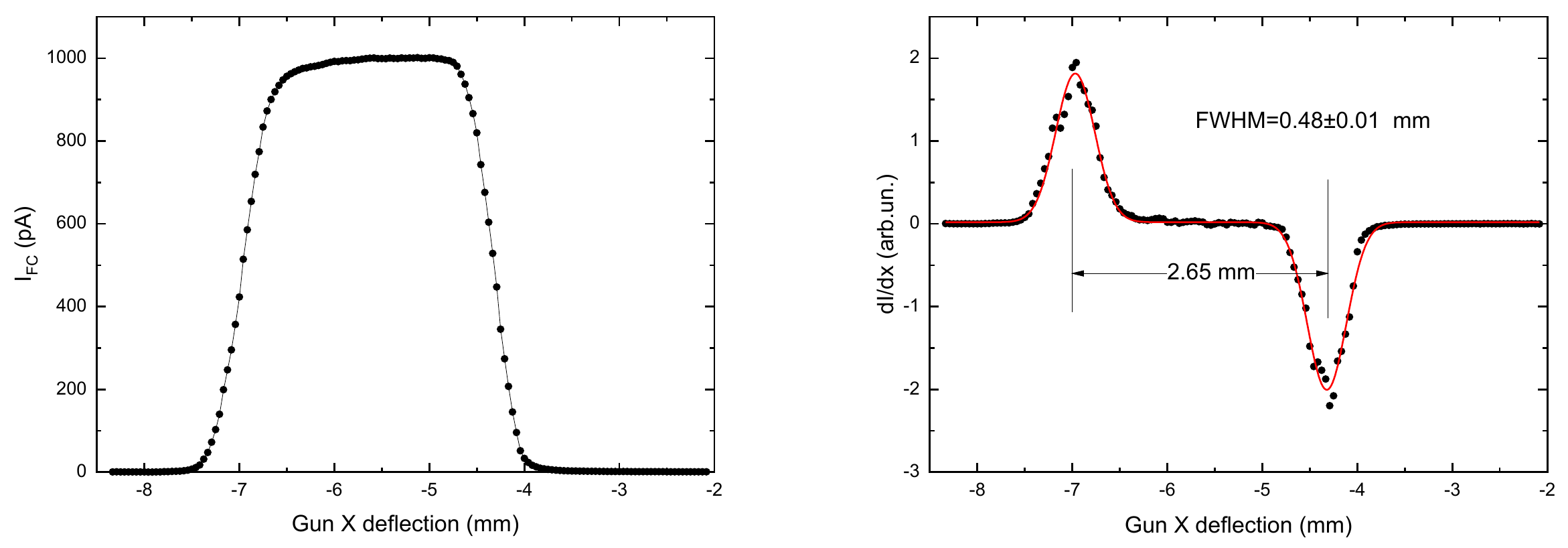}
 \caption{Measurement of the Faraday cup current (left) and its derivative (right) when sweeping the electron beam across the diameter of the cup.
  \label{fig:FC_scan}}
\end{figure}

The beam profile is measured {\em in situ} with the Faraday cup. A beam sweep is performed across the front face of the cup, along its diameter. The current of the cup is measured and a typical current profile can be seen in Fig.~\ref{fig:FC_scan}~(left), which was obtained for a beam with energy $E_e = 90$~eV and current of approximately $I_{gun} = 1$~nA: as can be seen the measured current is zero when the gun is shooting outside of the central hole of the cup (for $x < -7.5$~mm and $x > -4$~mm), while it is different from zero when the beam is shooting inside the cup (for $-7.5 < x < -4$~mm). 

By taking the derivative of the current profile across the cup the beam profile can be measured. This is shown in Fig.~\ref{fig:FC_scan}~(right): the resulting structures are fitted with two Gaussian distributions, and their width~$\sigma$ (which is constrained in the fit to be the same for both Gaussians) is taken as a measurement of the beam profile width. The result for the shown beam configuration is~$\sigma = (0.48\pm0.01)$~mm, which is a typical value for all electron energies and beam currents investigated in this paper.  

\begin{figure}[tb]
  \centering
\includegraphics[width=0.49\textwidth]{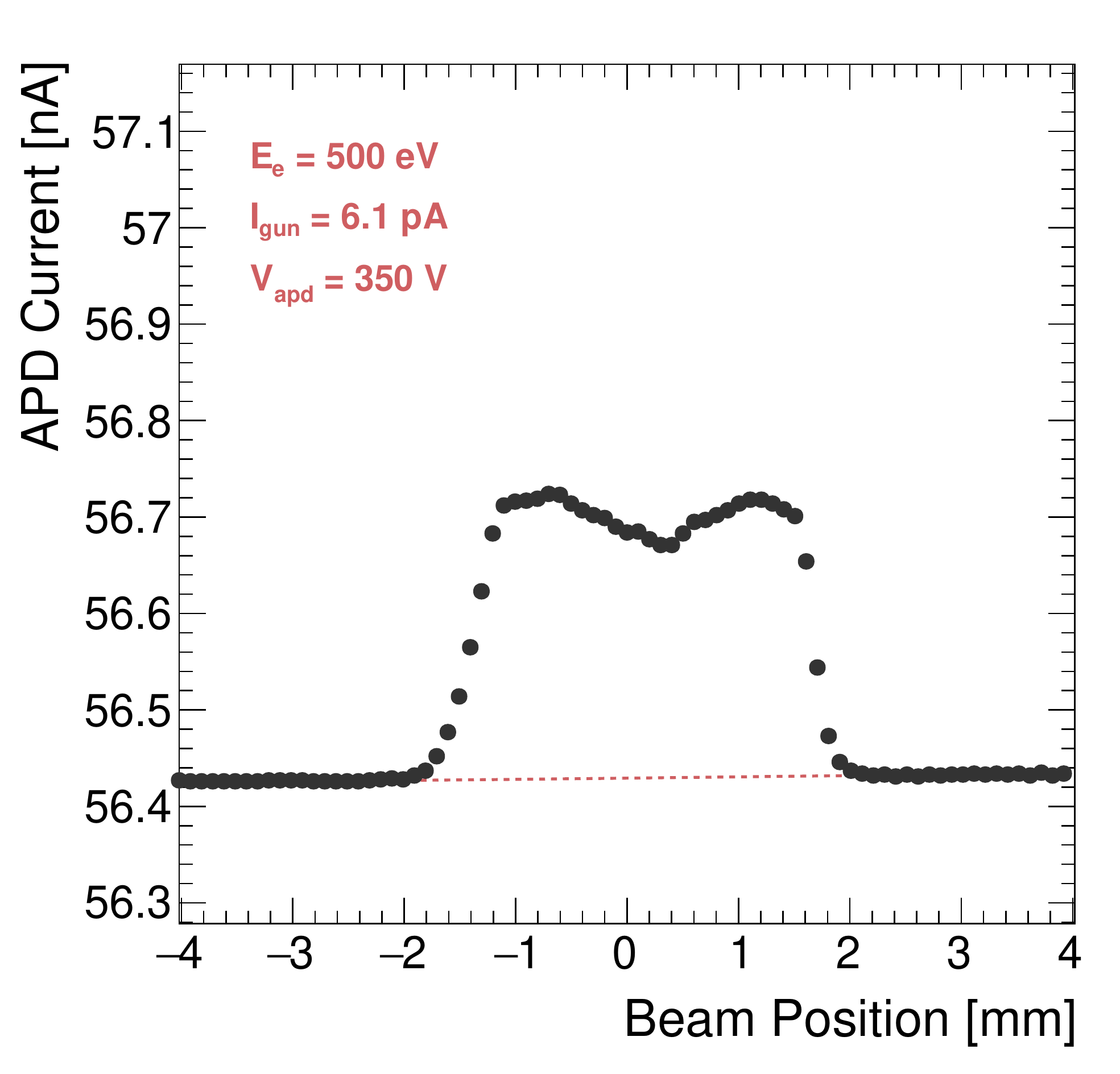}
\includegraphics[width=0.49\textwidth]{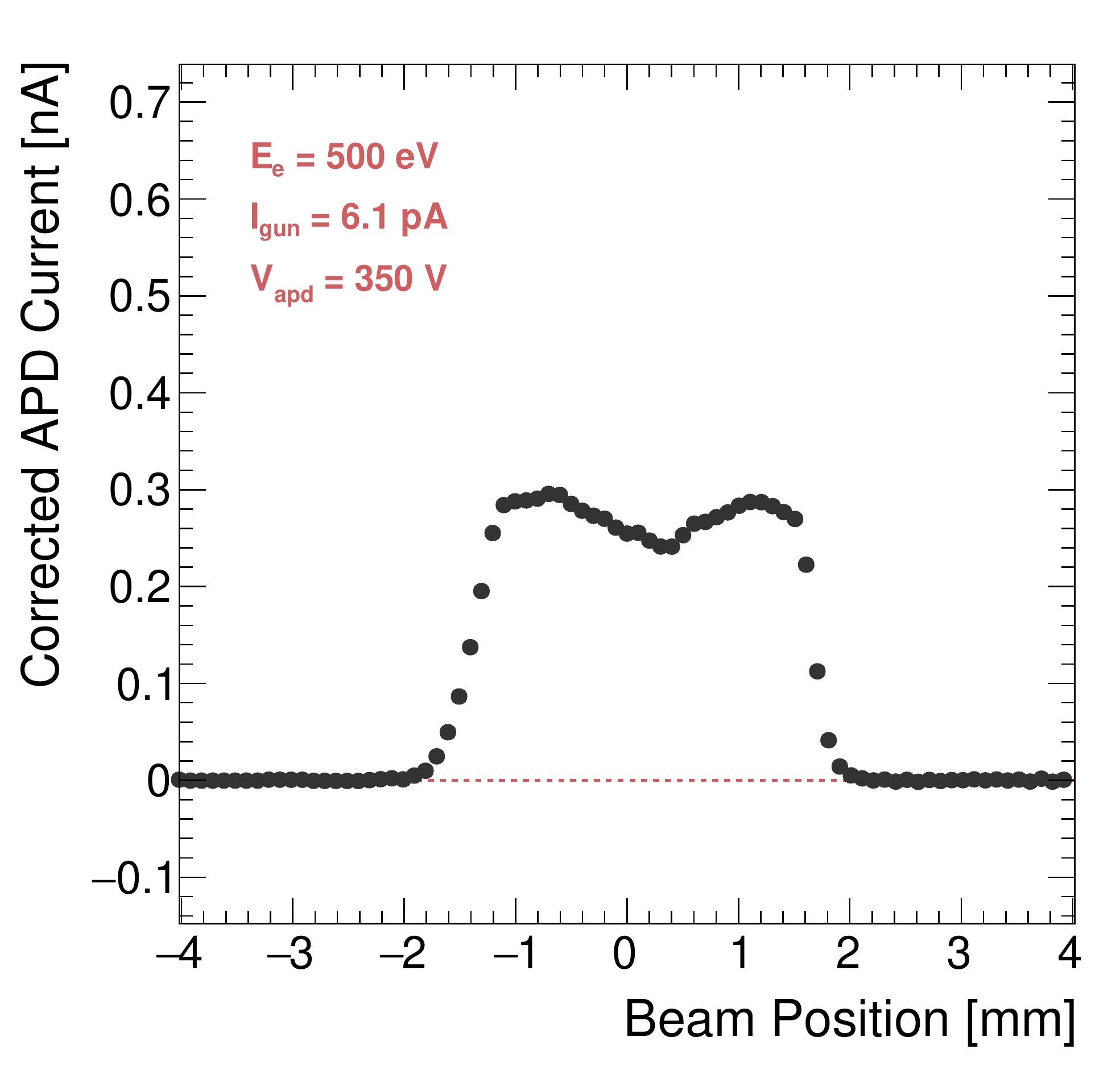}
 \caption{Typical measurement of the APD current when sweeping the electron beam across its sensitive area, along its diameter, before~(left) and after~(right) the dark current subtraction. This sweep was performed with the values of electron energy $E_e$ and gun current $I_{gun}$ reported in the plots.
  \label{fig:apd_scan}}
\end{figure}

A similar technique is used to measure the response of the APD. A beam sweep is performed across the sensitive area of the photo-diode, and the current generated by the APD is recorded. A typical scan is shown in Figure~\ref{fig:apd_scan}~(left): as can be seen the APD generates a non-zero (dark) current even when the gun is directed outside of the silicon. As soon as the gun enters the sensitive area of the diode, the measured current increases. As can be seen, the region in which this happens has an extension of about 3~mm, which coincides with the diameter of the Hamamatsu photo-diode.

To have a precise estimate of the current generated in the APD by the gun electrons, we subtract the contribution of the dark current. This is done by fitting the first 18 and last 16 points of the scan, which are sufficiently far from the sensitive region, with a third-degree polynomial function, so as to be able to describe possible temperature-related drifts during the scans (which take a few minutes). The dark current contribution is then subtracted, and the resulting background-subtracted scan is shown in Figure~\ref{fig:apd_scan}~(right).


\begin{figure}[tb]
  \centering
\includegraphics[width=0.49\textwidth]{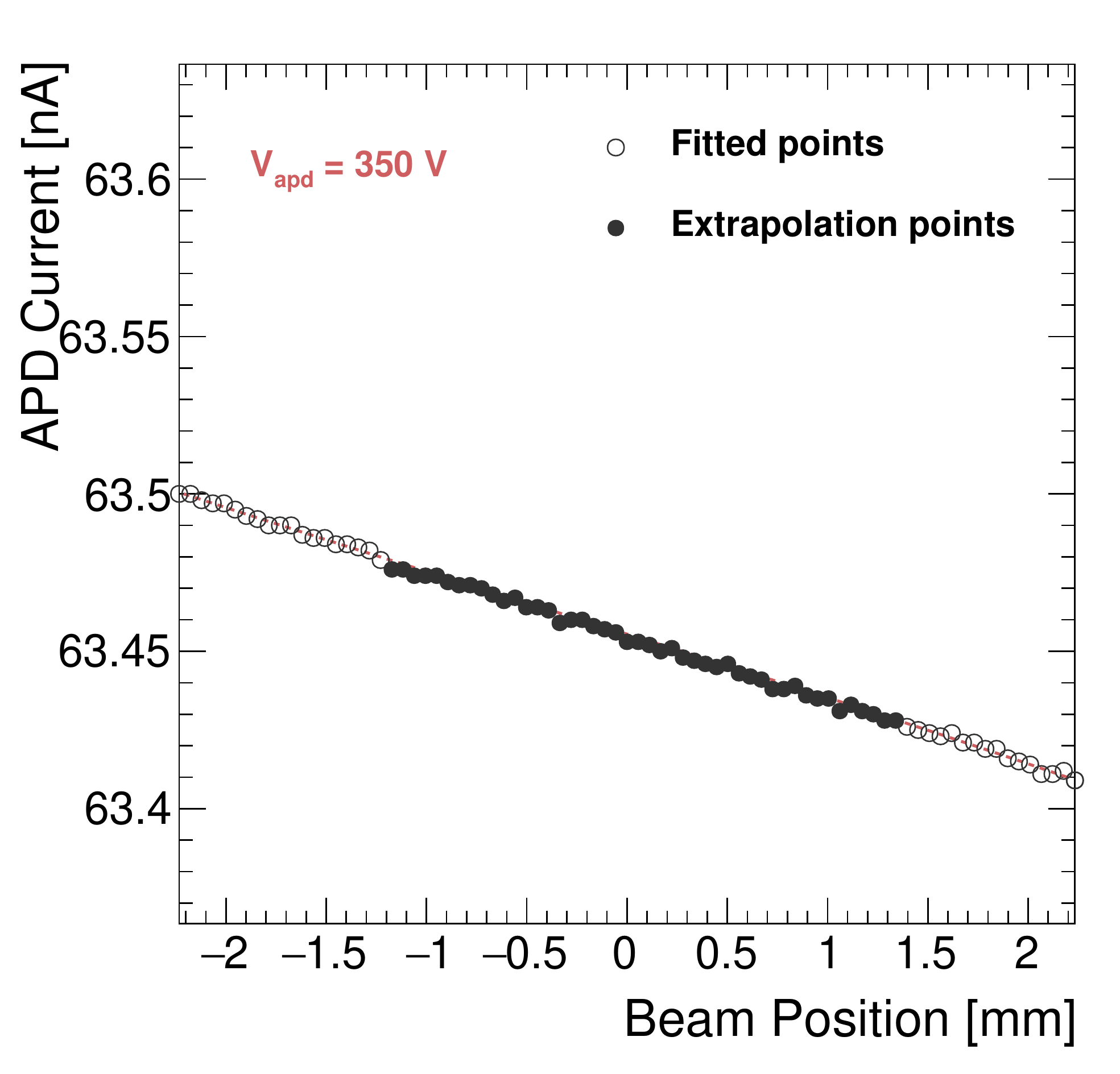}
 \caption{Example of empty APD scan used for the estimation of the systematic uncertainty connected with the fitting procedure. The open markers indicate the points used in the fit, and the red line the result of the polynomial fit. The fit result is compared to the markers in the central region of the scan (solid black) to evaluate the goodness of the extrapolation.
  \label{fig:apd_syst}}
\end{figure}

Possible systematic uncertainties on the dark-current fitting method have been estimated by taking `empty' scans, i.e.~scans in which the gun was never directed over the sensitive region of the APD. An example of such scan is shown in Figure~\ref{fig:apd_syst}, where one can see the APD dark current drifting during the course of the scan. The same fitting procedure is then applied to the empty scan, by fitting the first 18 and last 16 points of the scan (open markers in the figure), and then the fitted function is compared to the measured points in the rest of the scan (solid markers). The difference between prediction and measurement has an average compatible with zero, indicating that there is no bias in the procedure, while its width is found to be 1.22~pA. This value is taken as systematic uncertainty on the dark-current subtraction method. 


\begin{figure}[tb]
  \centering
\includegraphics[width=0.59\textwidth]{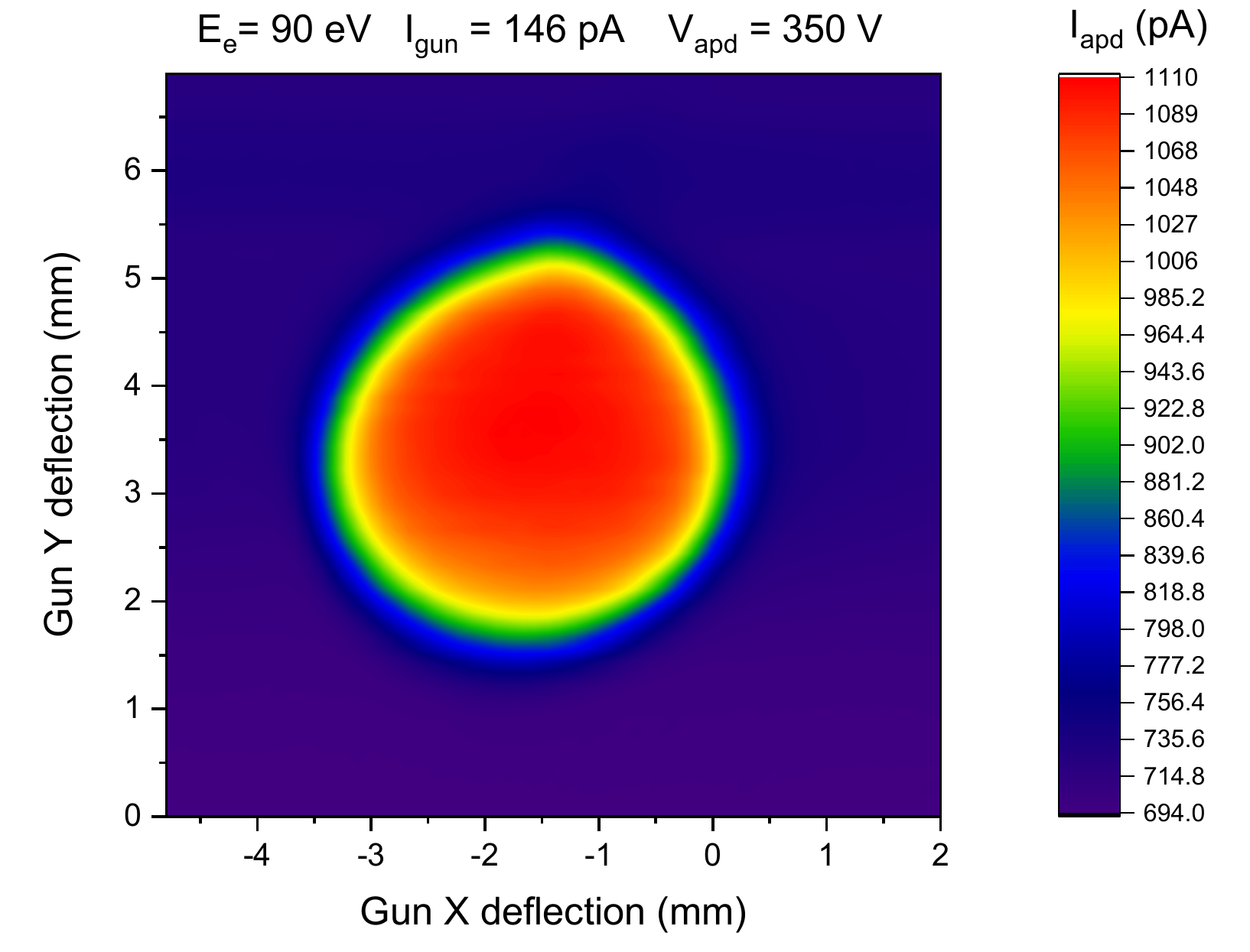}
 \caption{Measurement of the APD current as a function of the electron beam position, when sweeping the electron beam across the full front face of the APD. The scan was performed with 90~eV electrons and a gun current of $I_{gun} = 146$~pA.
  \label{fig:2d_scan}}
\end{figure}

By proceeding in a similar fashion, we also performed 2-dimensional scans, directing the electron beam across the whole surface of the APD. This is shown in Figure~\ref{fig:2d_scan}, where the measured APD current is shown on the $z$-axis, as a function of the $x-y$ position of the incident electron beam. The outline of the circular APD sensitive region is clearly visible.


\clearpage

\section{Results}

To measure the APD response to the gun current, we perform beam sweeps for multiple gun currents in the range $I_{gun} \in 0.1 \div 100$~pA. As the electron source is not bunched, this corresponds to electron rates between about 600~kHz and 60~MHz. The gun current is measured with the Faraday cup before and after the APD sweep. The uncertainty on $I_{gun}$ is taken as either half of the absolute difference between these two measurements, or 0.5\% of their average value, whichever is largest.

The APD current is extracted from the sweep profiles. Once the dark-current contribution is subtracted as described in the previous section, the maximum APD current $I_{apd}^{max}$ is identified, and only points $i$ with $I_{apd}^i > 0.5\cdot I_{apd}^{max}$ are considered. The APD current $I_{apd}$ is defined as the average of these points. The uncertainty on $I_{apd}$ is dominated by the systematic uncertainty on the background subtraction (1.22~pA), while the instrumental resolution~(0.1~pA) is found to be negligible. 

\begin{figure}[htb]
  \centering
  \includegraphics[width=0.49\textwidth]{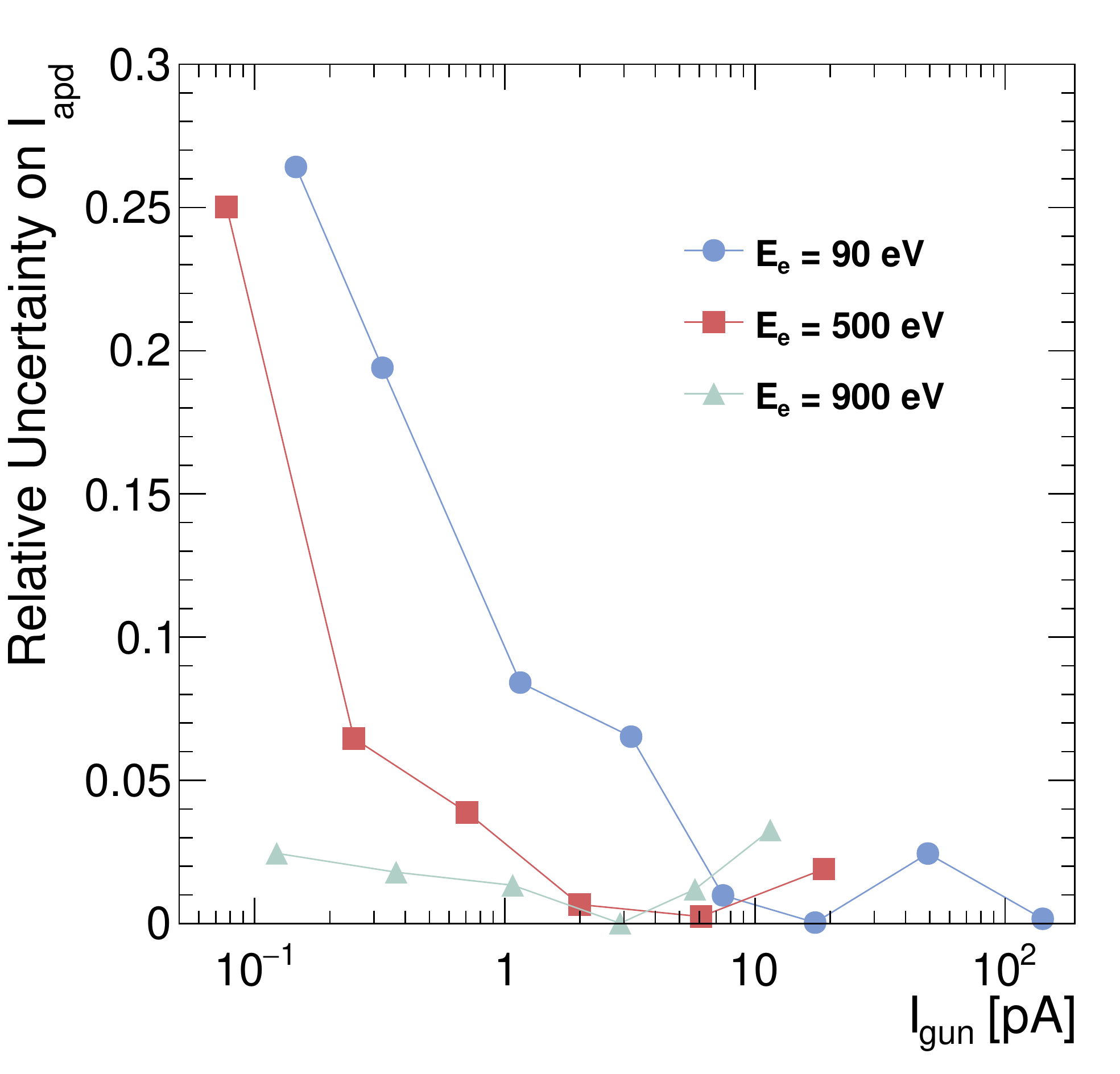}
 \caption{Relative uncertainty on the APD current definition, as a function of the gun current $I_{gun}$, for $E_e = 90$~eV~(blue circles), 500~eV~(red squares), and 900~eV~(green triangles).
  \label{fig:systCurrent}}
\end{figure}

We have estimated the systematic uncertainty connected to the choice of this definition of $I_{apd}$ by computing, in each point, an alternative current estimator. This alternative estimator defines $I_{apd}$ by first computing the integral of the background-subtracted scan, and then dividing the result by the APD nominal diameter~(3~mm). The two $I_{apd}$ definitions are found to be in good agreement,  except for the integral method giving values systematically 10\% larger than the nominal method, because of its complete inclusion of the tails. After taking into account this constant scale factor, we take residual differences between the two methods as a systematic uncertainty: this is summarized in Figure~\ref{fig:systCurrent}. As can be seen the relative uncertainty is typically no larger than a few percent, except for the low $I_{gun}$ values at $E_e = 90$ and 500~eV, where it rises up to about 25\%. 

\begin{figure}[p]
  \centering
  \includegraphics[width=0.59\textwidth]{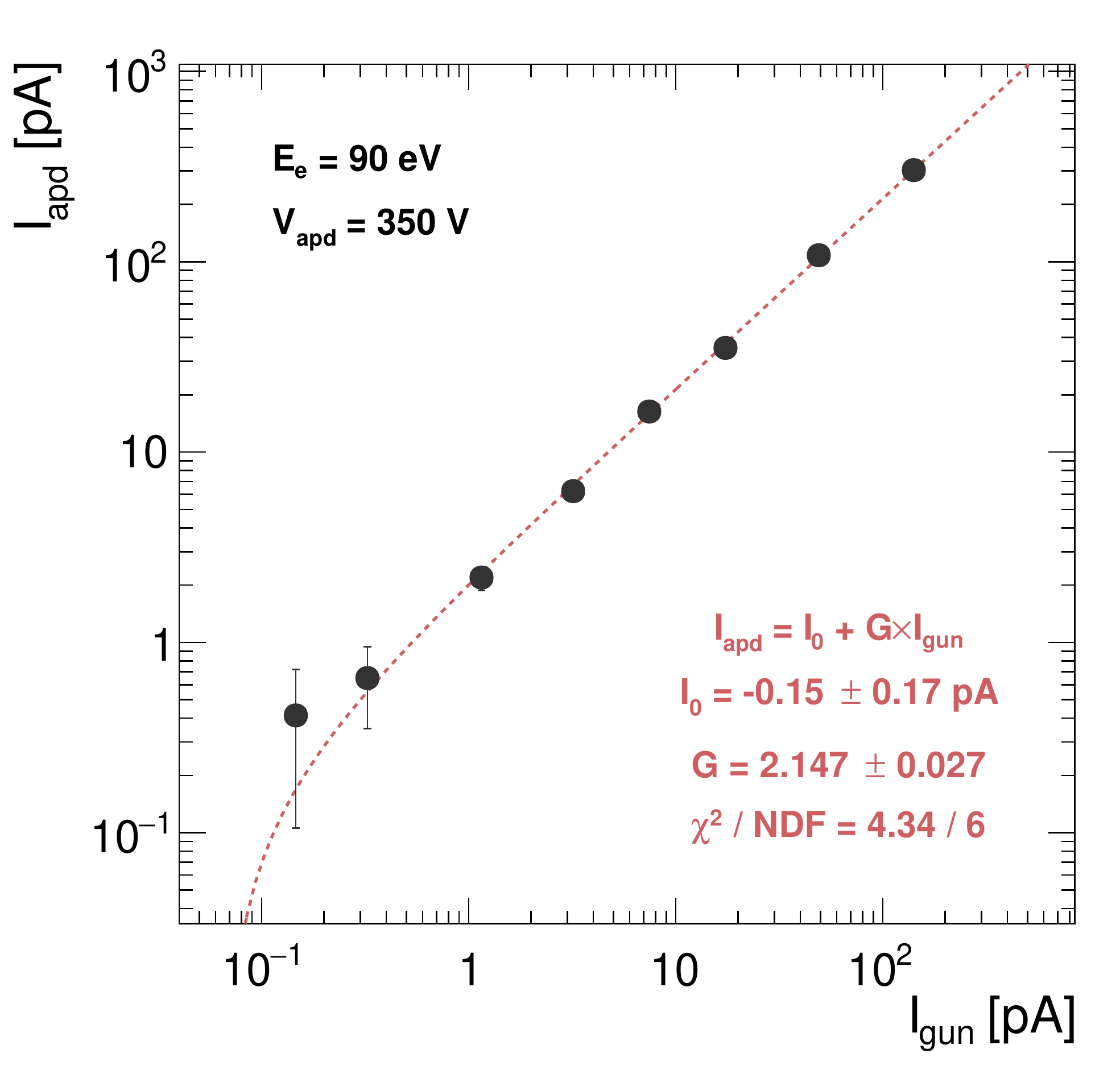}
  \includegraphics[width=0.59\textwidth]{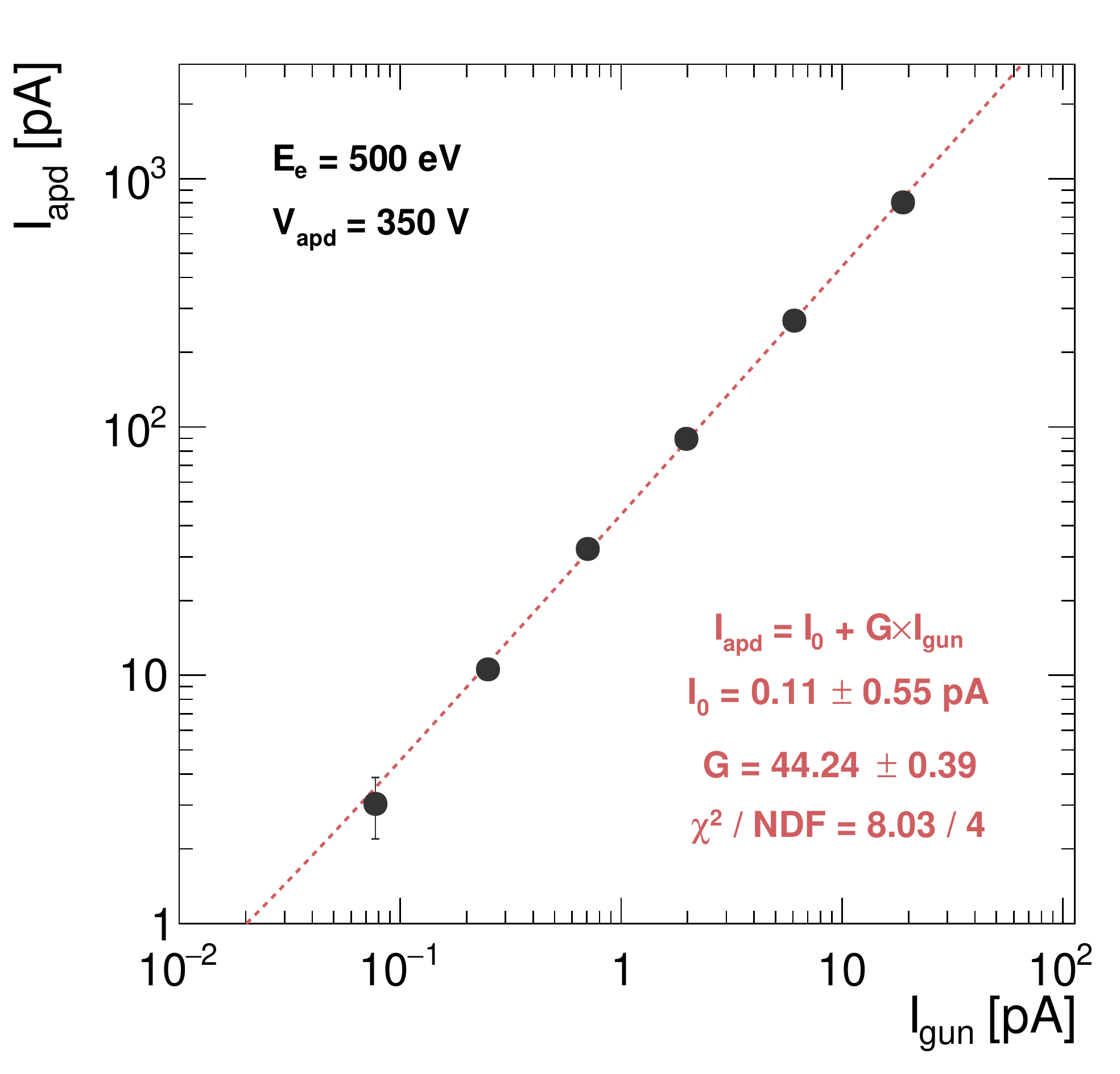}
  \includegraphics[width=0.59\textwidth]{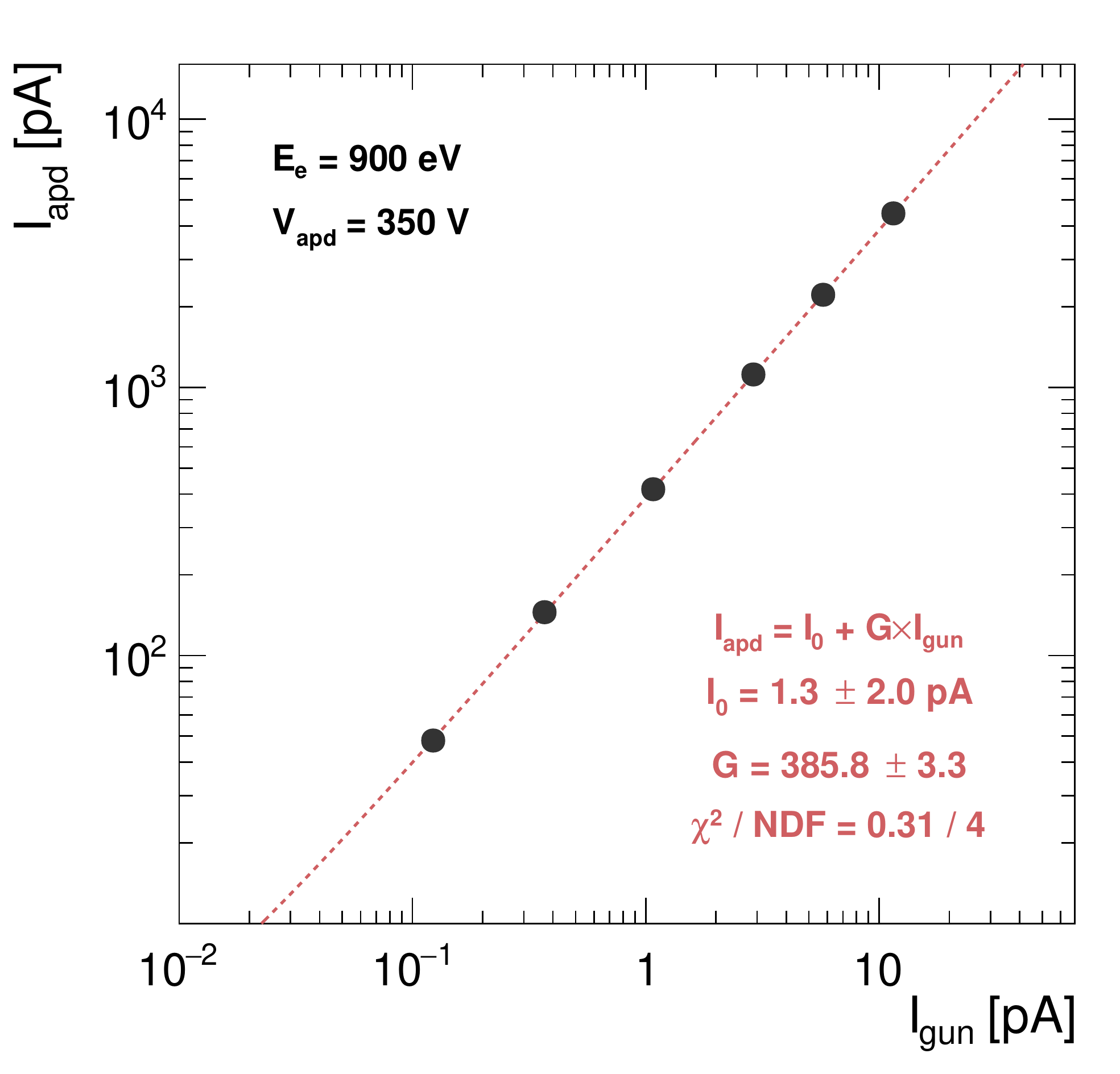}
 \caption{APD current as a function of the electron gun current, for $E_{e} = 90$~eV (top), 500~eV~(center) and 900~eV(bottom).
  \label{fig:i_vs_i}}
\end{figure}

The results of the APD characterization to electrons are shown in Figure~\ref{fig:i_vs_i}, for $E_{e} = 90$~eV (top), 500~eV~(center) and 900~eV (bottom). As can be seen a linear dependence between $I_{apd}$ and $I_{gun}$ is observed for all three energies, and the trends have been fitted with the function:
$$
I_{apd} = I_0 + G\cdot I_{gun},
$$
shown with a dashed red line in the plots. As can be seen by the $\chi^2$ values (reported in the figures) the linear hypothesis seems to be supported by the data, and in all cases the intercept $I_0$ is found to be compatible with zero. The fitted values of $G$ are taken as measurements of the effective APD gain, and are summarized in Table~\ref{tab:G}. The effective APD gain $G$ is found to increase with increasing electron energy~$E_e$. Two main factors determine the production of a current in the APD, and both of them depend on $E_e$: the probability that the electron is not absorbed in the inert layer, which depends on the mean free path of electrons in SiO$_2$, which has a non-linear dependence on $E_e$~\cite{meanfree}; and the number of of electron-hole pairs created in the active layer, which depends on the energy with which the electron enters such layer and on the energy needed to create an electron-hole pair, which for Silicon is about $\epsilon = 3.66$~eV~\cite{eh_pair}.

Making basic assumptions, one can write the following relationship between $I_{apd}$ and $I_{gun}$, for each electron energy $E_e$:
\begin{equation}
\label{eq:gain}
I_{apd} = I_{gun} \cdot \bar{E}(E_e, d)/\epsilon \cdot G_{\gamma}
\end{equation}
where $G_{\gamma}$ is the APD gain for photons and $\bar{E}(E_e, d)$, which depends (non-linearly) on the electron initial energy ($E_e$) and on the thickness of the SiO$_2$ layer ($d$), is the average energy of the electrons after losing energy in the SiO$_2$ layer. This average takes also into account electrons which are completely absorbed in the SiO$_2$ layer, which would therefore have $\bar{E}(E_e, d) = 0$. The APD gain for (visible) photons $G_{\gamma}$ at a given APD bias $V_{apd}$ was obtained according to the method described in \cite{apd_cms}. We have obtained $G_{\gamma}(V_{apd} = 350\mathrm{V}) = 42.0 \pm 0.1$.


Inverting Equation~\ref{eq:gain} one can obtain an expression for $\bar{E}(E_e, d)$:
$$
\bar{E}(E_e, d) = \epsilon \cdot G/G_{\gamma}
$$
where $G \equiv I_{apd}/I_{gun}$ is the effective gain as defined in the above and reported in the first column of Table~\ref{tab:G}. The obtained values of $\bar{E}(E_e, d)$ are reported in the last column of Table~\ref{tab:G}. It must be noted that these averages are taken on distributions which are expected to be highly non-symmetrical, and they are small as a consequence of the large fraction of electrons which are completely absorbed in the SiO$_2$ layer.



\begin{table}
\centering
\begin{tabular}{ccc}
\hline
$E_{e}$ [eV] & G & $\bar{E}(E_e, d)$ [eV] \\
\hline
90 & $2.147 \pm 0.027$ & 0.1\\
500 & $44.24 \pm 0.39$ & 3.7\\
900 & $385.8 \pm 3.3$ & 32.1\\
\hline
\end{tabular}
\caption{Effective APD gain $G$ for different incident electron energies $E_e$, and mean energy $\bar{E}(E_e, d)$ with which they enter the APD active layer.
\label{tab:G}}
\end{table}


\section{Conclusions}

We have reported on the characterization of windowless silicon APDs with electrons with energies of 90, 500 and 900~eV. The characterization has been performed with the monoenergetic electron gun of LASEC laboratories, in University of Roma Tre. We have found that the APD current is linearly proportional to the current of the electrons hitting its surface, and we have measured the APD effective gain, defined as the ratio between the APD current and the incident electron current, and found that it increases from $2.147 \pm 0.027$ (for $E_e = 90$~eV) to $385.8 \pm 3.3$~(for $E_e = 900$~eV), when operating the APD at a bias voltage of $V_{apd} = 350$~V. While our apparatus is sensitive to beams of electrons of these energies, with the current levels of noise it was not possible to observe single-electron signals. Nevertheless, this is the first time that silicon avalanche photo-diodes are employed to measure electrons with $E_e < 1$~keV.

\section*{Data availability statement}

The data that support the findings of this study are available from the corresponding author upon reasonable request.

\section*{Acknowledgements}

This project has received funding from the ATTRACT project funded by the EC under Grant Agreement 777222. Partial financial support by Universit\`a Roma Tre, ``Piano Straordinario della Ricerca 2015, azione n.3: Potenziamento dei laboratori di ricerca'' is greatly acknowledged.

\end{document}